# CNN-DST: ensemble deep learning based on Dempster-Shafer theory for vibration-based fault recognition

Vahid Yaghoubi[1,2], Liangliang Cheng[1,2], Wim Van Paepegem[1], Mathias Kersemans[1]

[1]Mechanics of Materials and Structures (MMS), Ghent University, Technologiepark 46, B-9052 Zwijnaarde, Belgium.

[2]SIM M3 program, Technologiepark 48, B-9052 Zwijnaarde, Belgium.

**Abstract**

Nowadays, using vibration data in conjunction with pattern recognition methods is one of the most common fault detection strategies for structures. However, their performances depend on the features extracted from vibration data, the features selected to train the classifier, and the classifier used for pattern recognition. Deep learning facilitates the fault detection procedure by automating the feature extraction and selection, and classification procedure. Though, deep learning approaches have challenges in designing its structure and tuning its hyperparameters, which may result in a low generalization capability. Therefore, this study proposes an ensemble deep learning framework based on a convolutional neural network (CNN) and Dempster-Shafer theory (DST), called CNN-DST. In this framework, several CNNs with the proposed structure are first trained, and then, the outputs of the CNNs selected by the proposed technique are combined by using an improved DST-based method. To validate the proposed CNN-DST framework, it is applied to an experimental dataset created by the broadband vibrational responses of polycrystalline Nickel alloy first-stage turbine blades with different types and severities of damage. Through statistical analysis, it is shown that the proposed CNN-DST framework classifies the turbine blades with an average prediction accuracy of 97.19%. The proposed CNN-DST framework is benchmarked with other state-of-the-art classification methods, demonstrating its high performance. The robustness of the proposed CNN-DST framework with respect to measurement noise is investigated, showing its high noise-resistance. Further, bandwidth analysis reveals that most of the required information for detecting faulty samples is available in a small frequency range.

# 1   Introduction

Over years, the vibration-based fault detection methods have been utilized for a variety of applications, and thus, evolved from traditional methods to machine learning (ML)-based methods [1]. Although ML-based methods have become very popular among researchers [1], their performances are dependant on not only the quality of vibration data but also the features extracted from the vibration data, the features selected to train the classifiers, and the classification methodology used for pattern recognition. Therefore, to improve the performance of these ML-based methods, one could either use proper feature selection techniques [2,3] or employ suitable pattern recognition methods [4]. However, it is well-known that these methods cannot give high accuracies for all applications and datasets, depending on the level of outliers, noise, errors, nonlinearities, and data redundancy present in the data [5]. To treat this problem, ensemble methods have been suggested in order to suppress the weaknesses and to boost the strengths of each classifier [6–8]. This could lead to higher accuracy and lower variance compared to the classification by the individual models.

Deep learning (DL) is a new trend among researchers in the domain of condition monitoring of machinery due to its promising result and automated feature learning [9–16]. An extensive review of using deep learning for machine health monitoring is presented in [10]. However, since tuning the hyperparameters is burdensome, acquiring an accurate and robust model is very challenging [17,18]. Further, the individual model sometimes has low generalization capability [19]. Therefore, ensembles of deep learning methods for fault detection have been recently employed. In [17], Ma and Chu proposed to create an ensemble by using different deep learning methods, e.g. Convolution Residual Network, Deep Belief Network, and Deep Autoencoders, for fault diagnosis. These models have been then combined by using multi-objective optimization. Chakraborty et. al. used stacked autoencoders to extract features and then use probabilistic neural networks to create an ensemble to be used for outlier detection [18]. The models have been combined by using a majority voting. In [19], Li et. al. created an ensemble model by using different variations of deep autoencoders (DAE) as the base models for fault diagnosis. A weighted voting method has been devised for fusion. Zhang et. al. proposed to use non-negative sparse constrained deep neural networks as the base model for fault diagnosis [20]. The models are then combined by using the Dempster-Shafer theory of evidence. Chen and Jahanshahi developed a new convolutional neural network (CNN) architecture to detect cracks and then employed the Naïve Bayes method for data fusion [21].

To employ an ensemble of deep learning models for any application, two main questions should be addressed:  i) how to impose diversity to the models in the ensemble, and ii) how to combine their outputs to make a final decision.

The first issue has been investigated in the literature by employing different classification methods, different types of features, different training samples, etc. [22]. In this regard, two widely used approaches are bootstrap aggregation [23], and adaptive boosting [24]. The common practice is to generate a pool of classifiers and select some of them to assure accuracy and diversity in the ensemble [25–28]. In [25], Yang proposed a selection technique based on accuracy and diversity (SAD) in which diversity has been evaluated based on Q-statistics. Faria et. al. [26], proposed a rank aggregation technique based on pairwise accuracy and diversity of the classifiers. To evaluate the diversity, a combination of several criteria has been used. Recently, Graph-theory has been employed to select proper classifiers from a pool [27,28]. Also, attempts have been made to choose proper classifiers by using information-related criteria [29,30], although this is not well established yet.

To tackle the second question, various methods have been developed. From basic operations [8] to more advanced forms like multilayered perceptrons [31], Bayes combination [32], fuzzy integrals [33], Dempster-Shafer theory (DST) of evidence [34]. Due to the proven advantages of the DST-based methods over other combination methods [5,22,35], it is selected as the combination method in this study. For instance, in [36–38] it is shown that the DST-based ensemble model could outperform the best individual model. However, when conflicting evidences come from different models, it could lead to counter-intuitive results. To treat this issue, one could employ different preprocessing techniques on the evidences. In this regard, Deng et. al. employed information from the confusion matrix to improve the basic belief assignment (BBA) [39]. In [40], Qian et. al. proposed to employ Shanon's information entropy together with Fuzzy preference relations (FPR) to reduce the conflict between the evidences. Xiao proposed to adjust the distance-based support degree of the evidences (SD) by utilizing the belief entropy and FPR [41]. This has been then used as the weight for evidences before applying Dempster's rule of combination. In [42], a criterion based on BBA similarity and belief entropy was used to weigh the evidences before applying Dempster's rule of combination. In [43], a $k$-nearest neighbor algorithm was employed to evaluate the BBAs. Wang et al. [44] proposed to employ a linear combination of two weightings, one based on the credibility of the evidences and the other based on the support degree of the evidences with respect to the focal elements with the largest BBAs. Zhunga et. al. developed a reliability-based technique to transform the classifiers' outputs to the BBAs [45,46].

This study proposes an ensemble deep learning framework for vibration-based fault recognition of complex metal parts. The framework is based on Convolutional Neural Network (CNN) and Dempster-Shafer theory (DST) of evidence, and is called CNN-DST. In this regard, the contribution of this study is threefold:

(i) A CNN architecture to detect defected samples with reasonably good accuracy
(ii) A new classifier ranking/selection procedure to create the ensemble

(iii) An improved DST method to perform the fusion over the ensemble

The paper is organized as follows. In Section 2, basic concepts of the Dempster-Shafer theory are presented. In Section 3, different steps of the proposed CNN-DST framework are elaborated. In Section 0, the proposed CNN-DST framework is applied to an experimental dataset generated from broadband vibration response of first-stage turbine blades with complex geometry and different damage types and severities. In Section 5 concluding remarks are presented.

## 2 Dempster-Shafer theory of evidence

Dempster-Shafer theory (DST) [34], is a key method in the evidence theory. In this paper, it is used for information fusion at the decision level. The basic concepts of DST are provided here.

### 2.1 Basic concepts of DST

**Definition 1**. Frame of discernment

Let $\mathfrak{E} = \{E_1, E_2, \ldots, E_K\}$ be a set of finite numbers of mutually exclusive and collectively exhaustive events $E$, then it is called the frame of discernment (FOD), $2^{\mathfrak{E}}$ is its power set, and $\mathcal{A}$ is called proposition provided that $\mathcal{A} \in 2^{\mathfrak{E}}$.

**Definition 2**. Basic Belief Assignment (BBA)

BBA maps each proposition $\mathcal{A} \subseteq \mathfrak{E}$ into the bounded range [0, 1] provided that

$$\begin{cases} m(\emptyset) = 0 \\ \sum_{\mathcal{A} \in 2^{\mathfrak{E}}} m(\mathcal{A}) = 1 \end{cases} \quad (1)$$

If $\mathcal{A} \subseteq \mathfrak{E}$ and $m(\mathcal{A}) \neq 0$, then $\mathcal{A}$ is called a focal element.

**Definition 3**. Dempster's rule of combination

For two independent bodies of evidence (BOEs) characterized by the BBAs $m_1$ and $m_2$ in the FOD $\mathfrak{E}$, the Dempster's rule of combination is defined as

$$m(\mathcal{A}) = (m_1 \oplus m_2)(\mathcal{A}) = \begin{cases} \dfrac{1}{1-K} \sum_{\mathcal{C}_1 \cap \mathcal{C}_2 = \mathcal{A}} m_1(\mathcal{C}_1) m_2(\mathcal{C}_2) & \mathcal{A} \subset \mathfrak{E}, \mathcal{A} \neq \emptyset \\ 0 & \mathcal{A} = \emptyset \end{cases} \quad (2)$$

$K$ is a measure of conflict between the two BOEs. It is defined as,

$$K = \sum_{\mathcal{C}_1 \cap \mathcal{C}_2 = \emptyset} m_1(\mathcal{C}_1) m_2(\mathcal{C}_2), \quad (3)$$

The combination rule can be applied to two BBAs if $K < 1$, otherwise the combination may give counterintuitive results.

## 2.2 Conflict measures

For proper DST combinations, two conditions should be satisfied, 1) $K \neq 1$, and 2) the BOEs are independent of each other. To determine the conflict between the BOEs, several measures have been defined. In the following, some of them are introduced. In this regard, let $\mathcal{A}_i$ be a proposition with cardinality $|\mathcal{A}_i|$ of a BOE characterized by BBA $m$.

### 2.2.1 Belief entropy

Belief entropy estimates the available uncertainty in the form of BBA. Deng entropy[47] is one of the common methods which is defined as,

$$E_d = -\Sigma_i m(A_i) \log\left(\frac{m(A_i)}{2^{|A_i|} - 1}\right) \quad (4)$$

In this definition, larger cardinality results in larger Deng entropy, which means more information. large Deng entropy for one evidence could be interpreted as more support from the other evidences.

### 2.2.2 Belief divergence

Belief divergence indicates the difference between two BBAs. Belief Jensen–Shannon (BJS) is a recent divergence measure proposed by Xiao [48]. For two BBAs $m_1$ and $m_2$ with $N$ propositions in the $\mathfrak{E}$, it is defined as,

$$BJS(m_1, m_2) = \frac{1}{2}\left[\Sigma_i m_1(A_i) \log\left(\frac{2m_1(A_i)}{m_1(A_i) + m_2(A_i)}\right) + \Sigma_i m_2(A_i) \log\left(\frac{2m_2(A_i)}{m_1(A_i) + m_2(A_i)}\right)\right] \quad (5)$$

### 2.2.3 Disagreement degree

Disagreement degree indicates how much a BOE is outlier for the other BOEs [49]. Let,

$$\bar{m}(A) = \frac{1}{L}\sum_{j=1}^{L} m_j(A) \quad (6)$$

$$\bar{m}_{\sim q}(A) = \frac{1}{L} \sum_{j=1, j \neq q}^{L} m_j(A) \tag{7}$$

Then, the average distance to these two centers are,

$$SW = \frac{1}{L} \sum_{j=1}^{L} d_J(m_j, \bar{m}), \tag{8}$$

$$SW_{\sim q} = \frac{1}{L-1} \sum_{j=1, j \neq 1}^{L} d_J(m_j, \bar{m}_{\sim q}) \tag{9}$$

in which $d_J(m_1, m_2) = \sqrt{(m_1 - m_2)^T \mathbf{Jac}(m_1 - m_2)}$ is the Jousselme's distance with $\mathbf{Jac}(A, B) = \frac{|A \cap B|}{|A \cup B|}$ as the Jaccard's weighting matrix. The disagreement degree is thus defined as,

$$m^*_{\Delta E, \sim q} = 0.5 + \frac{1}{\pi} \arctan\left(\frac{SW - SW_{\sim q}}{\sigma}\right) \tag{10}$$

in which $\sigma$ is a parameter to adjust the effect of the $SW - SW_{\sim q}$ on the degree of disagreement. Here it is set to 2.

## 3  Proposed methodology: CNN-DST

Figure 1 shows the building block of the proposed method. It consists of four main stages:

I) Data collection: this is carried out with one actuator and two sensors to collect the magnitude frequency response function (|FRF|) of the samples for a wide frequency range.

II) CNN generation: this step is devised to make several CNN models. This is elaborated in Section 3.1.

III) CNN selection: in this step, the CNNs are ranked based on their performance on the validation dataset. This is explained in Section 3.2.

IV) CNN fusion: in this step, the outputs of the CNN ensemble are combined to improve the classification accuracy. This is discussed in Section 3.3.

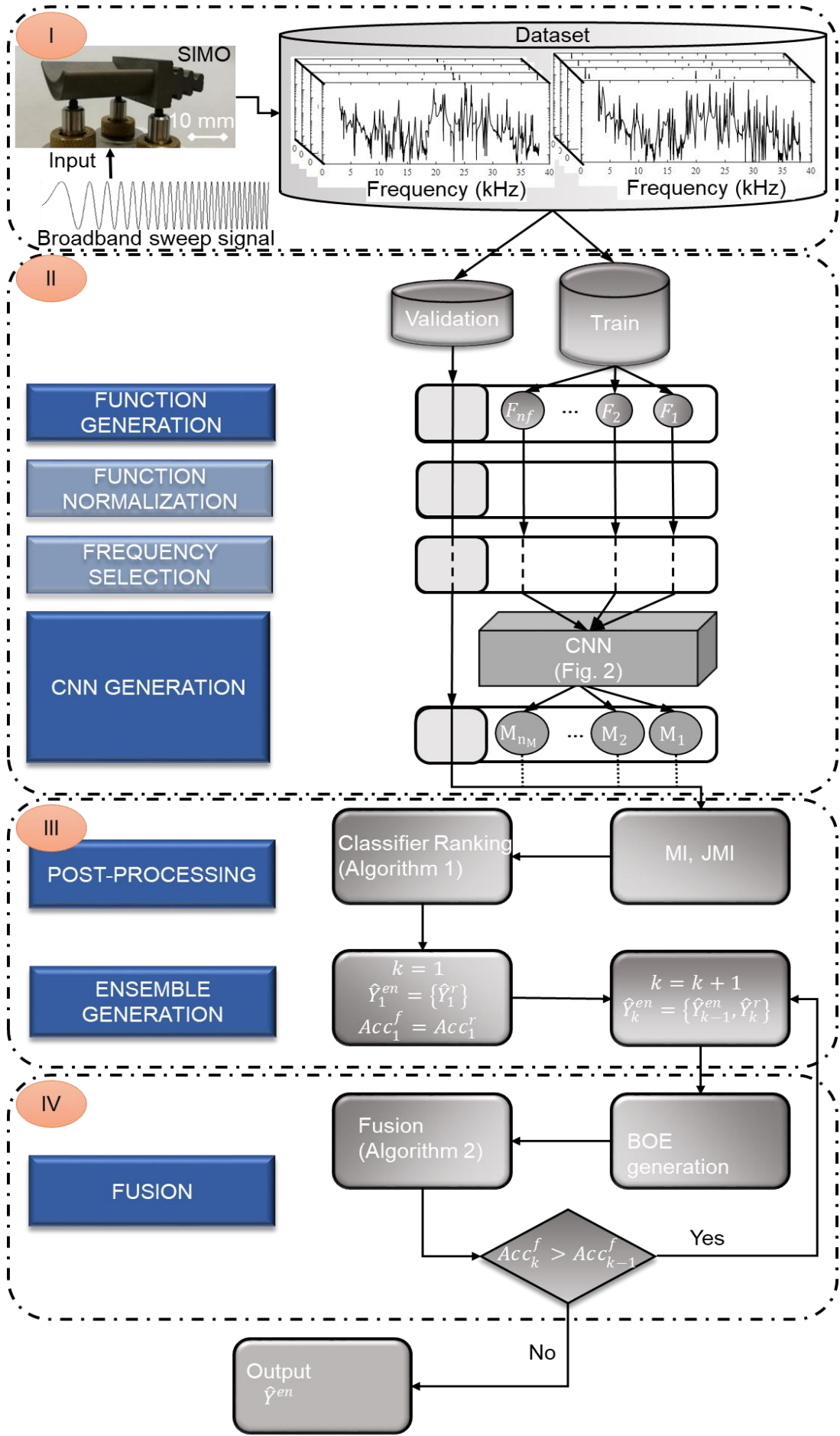

Figure 1. The procedure of the proposed ensemble deep learning framework CNN-DST

## 3.1 CNN-based classifier generation

At this stage, several models should be generated to create a pool of classifiers. For this purpose, a 4-step approach based on CNN is proposed: 1) Function generation, 2) Function normalization, 3) Data reduction/selection, and 4) CNN generation.

### 3.1.1 Function generation and normalization

Having a pool of classifiers is an inevitable task in ensemble learning. For this purpose, a variety of techniques have been proposed in the literature [38,50,51]. In this study, it is proposed to employ generative functions to extract information in the FRFs at different scales. That is, the FRFs are first passed through several generative functions and then used to train the CNNs. This could make the available information in the FRFs more accessible that could lead to faster convergence of the model. Besides that, it could bring diversity to the pool which is necessary to make a suitable ensemble model. In the current study the following functions have been used:

*Summation* and element-wise *multiplication* of the FRFs: to amplify the peaks and diminish the valleys. This, thus, increase the dynamic range of the FRFs and disclose the information at the high-amplitude frequencies.

*Logarithm* of the FRFs: this reduces the dynamic range of the FRFs. On the other hand, this reveals the information content at the low-amplitude frequencies.

As the next step, all the input signals are normalized to be in the range [-1, 1].

### 3.1.2 Data reduction/selection by LASSO

Normally, the FRFs are measured at many frequencies. Using all those frequencies to train the CNN models takes a lot of time and memory. Therefore, an additional step is devised here to select only the frequencies containing important information for performing classification. This could reduce the size of the FRFs which in turn, could reduce the required computational time and memory. This is achieved by employing the least absolute shrinkage and selection operator (LASSO) with the Least Angle Regression (LARS) method [52].

The LASSO method estimates the coefficient of a linear regression model under the constraint of having the sum of the absolute values of the coefficient less than a threshold. It is known for performing two tasks: regularization and variable selection. Recently, it attracts attention to be used as a feature selection method for classification [53]. Its procedure is as follows.

Let $X \in \mathbb{R}^{n_s \times n_f}$ and $Y \in \mathbb{R}^{n_s \times 1}$ be the matrices of input and output respectively. Here, $n_s$ and $n_f$ are the number of samples and features. The purpose is to find a linear regression model to fit the data, that is, $y_i = \beta_0 + \sum_{j=1}^{n_f} x_{ij}\beta_j$. The LASSO estimate of the coefficients $\beta$ is,

$$\hat{\beta}^{lasso} = argmin\left(\sum_{i=1}^{n_s}\left(y_i - \beta_0 - \sum_{j=1}^{n_f} x_{ij}\beta_j\right)^2\right) + \lambda \sum_{j=1}^{n_f} |\beta_j| \tag{11}$$

here $\sum_{j=1}^{n_f} |\beta_j|$ is the penalty term, and $\lambda$ is a tuning parameter to control the strength of the penalty term. The features could be ranked based on the magnitude of their associated coefficients $\beta$. One could relax $\lambda$, i.e. $\lambda = 0$, to obtain all the important features. In this case, the number of selected features, i.e. the features $x_{ij}$ with nonzero coefficient $\beta_j$, is thus limited to $\min(n_s, n_f)$. In the current study, the amplitudes of the FRFs create the feature space therefore, the number of features $n_f$ is equal to the number of frequencies available in the FRFs. Since normally $n_f \gg n_s$, the number of selected frequencies is upper limited to $n_s$. However, it should be noticed that this is the case for each generative function. By employing several generative functions, one can include more important frequencies for training the CNNs.

### 3.1.3 CNN generation

Figure 2 presents the overall structure of the proposed CNN model. The input signal, i.e. the magnitude of the FRF measured at $N_f$ frequencies, passes through 16 layers, and finally, the softmax layer predicts whether the part (associated with the FRF) is healthy or defected. The configuration of the layers is listed in Table 1. The values assigned to the hyperparameters of this network and its associated optimization algorithm could be chosen randomly. The values in the following are just suggestions for being consistent throughout the paper.

In the first layer, a 2D convolution layer with 64 kernels and the shape of [1 × 3] is performed on the input signal. The padding size is set to [1 × 2] such that the output has the same size as the inputs. The 2D convolution layer is used here to provide the possibility of having a multi-feature CNN. It has been suggested in [16] that to extract important information from time and/or frequency domain vibration data for damage detection, one should use a large kernel size for the first convolution layer and then use small ones for the other convolution layers. However, in this study, the LARS technique is employed for rough selection of those frequencies containing crucial information for detecting damage. Therefore, one could simply employ a small kernel size for the first convolution layer too. This drastically reduces the number of unknown parameters, leading to faster convergence of the model. This is illustrated in Table 2, in which the models converged faster without losing their accuracies. The output of this layer will be normalized across each mini-batch to have zero mean and unit variance. This is obtained by passing the mini-batch data through the batch normalization layer. This layer acts as a regularizer that could speed up the training procedure and reduce the sensitivity of the network to the initial points. Two rectified linear units (reLU) and one tangent hyperbolic are devised as activation layers to reduce or remove the effect of the negative values. The Dropout layer randomly disconnects some of the connections. This could help to prevent overfitting of the model to the trained data. The hyperparameter of the Dropout layer was set to 0.2. The other parameters of the network were set to their default values in Matlab.

The first fully connected layer in the 15$^{th}$ layer serves as the flatten layer. This layer transforms the feature obtained from the 14$^{th}$ layer to the proper size to be used as the input to the subsequent layer of the network. The second fully connected layer serves as the final classifier. The softmax layer is devised here to predict the health status of the associated sample by giving it a score between zero and one.

The model is implemented using the Matlab 2019a, deep learning toolbox. For training, the adam optimizer default parameters are utilized. The batch size was 128, the maximum used epochs set to 200, and the initial learning rate was 0.005 which decayed by the factor of 0.005 every 10 epochs. The other parameters of the optimization algorithm were set to their default values in Matlab.

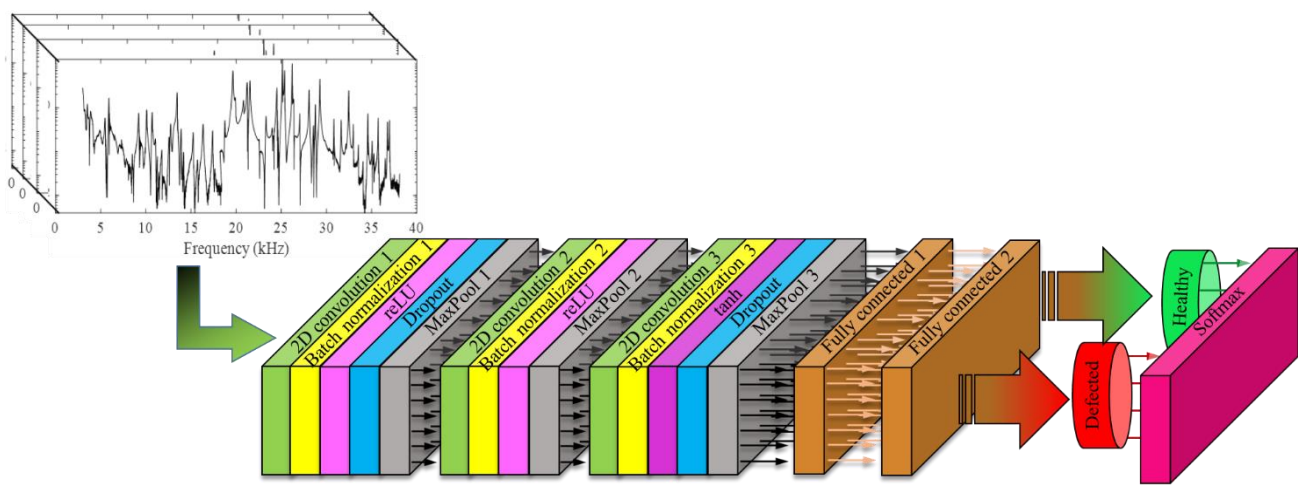

Figure 2. The proposed CNN structure

Table 1. Configuration of the layers in the proposed CNN

| Layer | Kernel Shape | Kernel # | Stride | Padding size | Output size |
|---|---|---|---|---|---|
| **Convolution 1** | $1 \times 3$ | 64 | 1 | $1 \times 2$ | $64 \times n_f$ |
| **MaxPool 1** | $1 \times 2$ | – | 1 | – | $64 \times \frac{1}{2} n_f$ |
| **Convolution 2** | $1 \times 3$ | 32 | 1 | $1 \times 2$ | $32 \times \frac{1}{2} n_f$ |
| **MaxPool 2** | $1 \times 2$ | – | 1 | – | $32 \times \frac{1}{4} n_f$ |
| **Convolution 3** | $1 \times 3$ | 16 | 1 | $1 \times 2$ | $16 \times \frac{1}{4} n_f$ |
| **MaxPool 3** | $1 \times 2$ | – | 1 | – | $16 \times \frac{1}{8} n_f$ |
| **FC 1** | 150 | – | – | – | 100 |
| **FC2** | 100 | – | – | – | 2 |

$n_f$ is the number of frequency lines used for training

## 3.2 CNN selection/ranking

Classifier ranking/selection is an inevitable step for creating a suitable classifier ensemble. Several criteria have been proposed in the literature[27,28,54]. In these approaches, classifiers are selected based on a trade-off between the accuracy and diversity of the classifiers in the ensemble. Unlike the accuracy, there is no consensus about a metric to measure the diversity hence, several metrics have been proposed [26]. Another category of the methods used for classifier selection is based on information theory [29,30]. Although this category of selection methods has not been investigated thoroughly as a classifier selection, they have been extensively developed for feature selection [26,55–58]. In the current study, an information-based classifier ranking methodology is adopted here which is based on Joint Mutual information and "maximum of the minimum"[56].

Let $\Gamma = (\gamma_1, \gamma_2, \ldots, \gamma_N), \Omega = (\omega_1, \omega_2, \ldots, \omega_N), \Lambda = (\lambda_1, \lambda_2, \ldots, \lambda_N)$ be discrete random variables. Entropy $H$ measures the information available in each of these variables. It is defined as

$$H(\Gamma) = -\Sigma_{i=1}^{N} p(\gamma_i) \log(p(\gamma_i)) \qquad (12)$$

in which $p(\gamma_i)$ is the probability mass function. Mutual information $(MI)$ is the amount of information shared between two random variables. It is defined as,

$$MI(\Gamma;\Lambda) = MI(\Lambda;\Gamma) = H(\Gamma) - H(\Gamma|\Lambda) \qquad (13)$$

Here, $H(\Gamma|\Lambda)$ is called conditional entropy and measures the amount of information left in $\Gamma$ after introducing $\Lambda$. It is defined as,

$$H(\Gamma|\Lambda) = -\Sigma_{j=1}^{M}\Sigma_{i=1}^{N} p(\gamma_i,\lambda_j)\log(p(\gamma_j|\lambda_i)) \qquad (14)$$

The Joint mutual information ($JMI$) is defined as:

$$JMI(\Gamma,\Omega;\Lambda) = JMI(\Gamma;\Lambda|\Omega) + MI(\Omega;\Lambda) \qquad (15)$$

$$JMI(\Gamma;\Lambda|\Omega) = H(\Gamma|\Lambda) - H(\Gamma|\Lambda,\Omega) \qquad (16)$$

Now, let $Y \in \mathbb{R}^{n_s \times n_c}$ be the matrix of output in which, $n_s$, and $n_c$ are the number of samples, and classes. Further, let $\widehat{Y}_i, i = 1,2,\ldots,N_c$ be the output predicted by the $i^{th}$ classifier. Then Algorithm 1 is used to rank the classifiers. After ranking the classifiers, the classifiers will be added to the ensemble one by one. At each step, the performance of the ensemble will be assessed and the best one will be selected.

---

Algorithm 1. The proposed classifier ranking approach

---

**Inputs:** $\widehat{Y} = \{\widehat{Y}_1, \widehat{Y}_2, \ldots, \widehat{Y}_{N_c}\}, \widehat{Y}^r = \{\emptyset\}, Y$

$\widehat{Y}_s = argmax_{\widehat{Y}_i \in \widehat{Y}}\left(MI(\widehat{Y}_i;Y)\right), \widehat{Y}^r \leftarrow \widehat{Y}_s, \widehat{Y} \leftarrow \widehat{Y}\backslash\widehat{Y}_s$

$\widehat{Y}_s = argmax_{\substack{\widehat{Y}_j \in \widehat{Y}^r \\ \widehat{Y}_i \in \widehat{Y}}}\left(JMI(\widehat{Y}_i;Y|\widehat{Y}_j)\right), \widehat{Y}^r \leftarrow \widehat{Y}^r \cup \widehat{Y}_s, \widehat{Y} \leftarrow \widehat{Y}\backslash\widehat{Y}_s$

**while** $\widehat{Y} \neq \{\emptyset\}$ **do**

$\widehat{Y}_s = argmax_{\widehat{Y}_i \in \widehat{Y}}\left(min_{\widehat{Y}_j \in \widehat{Y}^r}\left(JMI(\widehat{Y}_i,\widehat{Y}_j;Y)\right)\right), \widehat{Y}^r \leftarrow \widehat{Y}^r \cup \widehat{Y}_s, \widehat{Y} \leftarrow \widehat{Y}\backslash\widehat{Y}_s$

**Output:** $\widehat{Y}^r = \{\widehat{Y}_1^r, \widehat{Y}_2^r, \ldots, \widehat{Y}_{N_c}^r\}$

---

### 3.3 CNN fusion

In this section, the output of the CNNs in the ensemble will be combined to improve the classification performance. To this end, first, the BOEs should be made from the classifiers' outputs. They are explained in the following.

*3.3.1 Generating BOEs*

Let $\widehat{Y}_i = [\widehat{y}_{i,1}, \ldots \widehat{y}_{i,k}, \ldots, \widehat{y}_{i,n_c}] \in \mathbb{R}^{(n_s \times n_c)}$ with $n_s$ and $n_c$ as the number of samples and classes respectively, be the result of the CNN $C_i, i = 1,2,\ldots,N_c$, then the BOEs with BBA $m_i$ in the FOD $\mathfrak{E} = \{E_1, \ldots, E_k, \ldots, E_{n_c}\}$ can be obtained as

$$m_i(E_k) = \mathbf{w}_i \otimes \hat{\mathbf{y}}_{i,k} \tag{17}$$

$$m_i(\mathfrak{E}) = 1 - \sum_{k=1}^{n_c} \mathbf{w}_i \otimes \hat{\mathbf{y}}_{i,k} \tag{18}$$

in which $m_i(\mathfrak{E})$ is the ignorance associated with the classifier, $\otimes$ is the Kronecker product, and $\mathbf{w}_i \in \mathbb{R}^{1 \times n_c}$ is a weighting factor. Different weightings to generate BOEs have been presented in [59]. In the current paper, to generate the BOEs $\mathbf{w} = \mathbf{1}$ which means no weighting, has been considered.

### 3.3.2 Improved DST-based fusion method

In this section, a new method for combining classifiers in the ensemble is proposed. This is an improvement to the method proposed in [38]. In [38], the credibility of the evidences is used as a weight to modify them before the combination. In the current study, the support degree of the evidences with respect to the focal elements with the largest BBAs is also included in the weight, which could be beneficial in reducing the conflict [44].

let $\mathfrak{E} = \{E_1, \dots, E_k, \dots, E_{n_c}\}$ be the FOD in which the BOEs with BBA $m_i$, $i = 1, 2, \dots, N_c$ are defined. Then, the following steps should be taken to modify the BOEs prior to combinations.

Step 1. Average belief divergence based on BJS in Eq. (5)

$$aBJS_i = \frac{1}{N_c - 1} \sum_{\substack{j=1, \\ j \neq i}}^{N_c} BJS(m_i, m_j) \tag{19}$$

Step 2. Evaluate the disagreement degree $(m^*_{\Delta E, \sim i})$ in Eq. (10)

Step 3. Evaluate the Support degree as

$$SD_i = (aBJS_i \times m^*_{\Delta E, \sim i})^{-1} \tag{20}$$

Step 4. Normalize the support degree as

$$\widehat{SD} = \frac{SD}{\sum_{i=1}^{N_c} SD_i} \tag{21}$$

Step 5. Evaluate the credibility degree by using the Deng entropy $E_d$ in Eq. (4) as,

$$CD = \exp(E_d) \times \widehat{SD} \tag{22}$$

Step 6. Normalize the Credibility degree as

$$\widehat{CD} = \frac{CD}{\max(CD)} \tag{23}$$

Step 7. Evaluate the support degree of evidences to the chief focal element

7.1. find the chief focal element

$$E^{chief} = \arg\max(\bar{m}) \tag{24}$$

with $\bar{m} = \frac{1}{N_c}\Sigma_{i=1}^{N_c} m_i$

7.2. Evaluate the support degree w.r.t chief focal element

$$SD_i^{chief} = m_i(E^{chief}), i = 1, 2, \ldots, N_c \tag{25}$$

7.3 Normalize the chief support degree

$$\widehat{SD}^{chief} = \frac{SD^{chief}}{\max(SD^{chief})} \tag{26}$$

Step 8. Evaluate the weight as follows,

$$\boldsymbol{W} = \theta\widehat{\boldsymbol{CD}} + (1-\theta)\widehat{\boldsymbol{SD}}^{chief} \tag{27}$$

in which $\theta$ is called the DST parameter. It determines the contribution of each part to the weighting.

Step 9. Normalize the weight as

$$\widehat{\boldsymbol{W}} = \frac{\boldsymbol{W}}{\text{sum}(\boldsymbol{W})} \tag{28}$$

Step 10. Evaluate the weighted average evidences as

$$\boldsymbol{WAE} = \Sigma_{i=1}^{N_c} \widehat{\boldsymbol{W}}_i \times m_i \tag{29}$$

Step 11. Combination based on Dempster's rule Eq. (2)

$$\widetilde{Y} = \bigoplus_{i=1}^{N_c} \boldsymbol{WAE} \tag{30}$$

The proposed fusion method is summarized in Algorithm 2. To illustrate the fusion procedure, a toy example is provided in Appendix I.

Algorithm 2. The proposed classifier fusion framework

---

**Inputs:** $C^{en}$: ensemble CNN,
        $N_c^{en}$: Number of CNNs in the ensemble
        $W$: weighting factor
**for** $i = 1$ **to** $N_c^{en}$ **do**
    Evaluate $m_i$ using Eq. (17) and (18)
    Evaluate $aBJS_i$ using Eqs. (5) and (19)
    Evaluate $m^*_{\Delta E,\sim i}$ using Eq. (10)
    Evaluate $\widehat{SD}$ using Eqs. (20) and (21)
    Evaluate $\widehat{CD}$ using Eqs. (4), (22), and (23)
    Evaluate $\widehat{SD}^{chief}$ using Eqs. (24)- (26)
    Evaluate $\widehat{W}$ by assigning the parameter $\theta$ and using Eqs. (27) and (28)
    Evaluate $WAE$ using Eq. (29)
Combine the results $\widetilde{Y}$ using Eqs. (2), and (30)
**Output:** $\widetilde{Y}$

---

## 4 Application to first-stage turbine blades

In this section, the proposed CNN-DST framework is applied to vibrational response data from Equiax Polycrystalline Nickel alloy first-stage turbine blades with complex geometry. Two views of its CAD model are shown in Figure 3. The cooling channel in the middle can be seen in the transparent view. By using one actuator and two sensors, i.e. Single Input Multiple Output SIMO, the amplitude of the frequency response function (|FRF|) has been collected from each blade in the range of [3, 38] kHz at 11253 frequency lines, see Figure 4. To create a database 150 healthy and 79 defected blades have been measured. The damages in the blades range from microstructural changes due to over-temperature, airfoil cracking, inter-granular attack (corrosion), thin walls due to casting, to service wear.

To commence the procedure, the data was divided into 70% for training and 30% for validation. Since the number of defected blades is almost half of the healthy ones, some randomly selected defected samples were repeated such that an equal number of healthy and defected samples, i.e. 105 healthy and 105 defected blades, have been used for training the CNNs. Let $X_1$ and $X_2$ be respectively the two acquired |FRF|s. Six functions are used to generate new features: summation $(X_1 + X_2)$, element-wise

multiplication $(X_1 \circ X_2)$, square[1] $X_1^2, X_2^2$, and logarithm $\log(X_1), \log(X_2)$. By using each of the features, one CNN is trained that will be referred to as the CNN("*feature*"). Moreover, one CNN is trained by using all the eight features, this is referred to as the CNN(*all*).

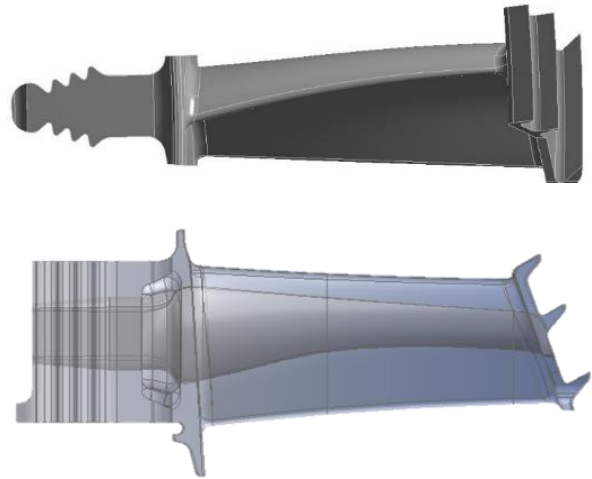

Figure 3. Two views of the CAD model of the Equiax Polycrystalline Nickel alloy first-stage turbine blade. Bottom plot shows a transparent view to illustrate the cooling channel.

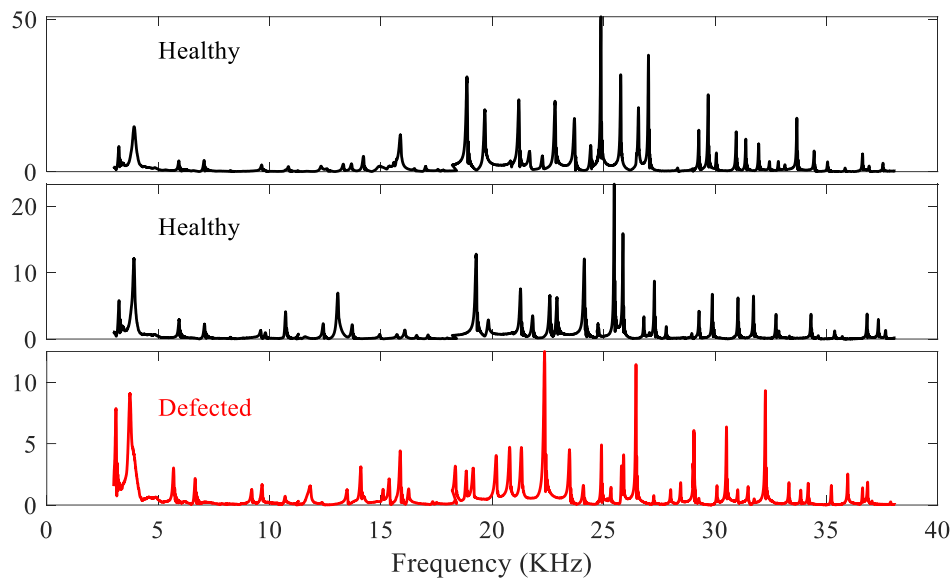

Figure 4. Three examples of the collected FRFs. Two top FRFs have been collected from healthy blades and the bottom FRF was from a defected blade.

In the step of frequency selection, the LARS algorithm has been applied to eight different functions of the FRFs. Since 210 training samples are available, from each function utmost 210 frequency lines have been selected. From the eight functions, in total 1360 unique frequency lines have been thus chosen for training.

---

[1] This can be considered as the element-wise multiplication of a vector by itself

In the step of combining the CNNs' outputs by using the proposed fusion method, the DST parameter ($\theta$) in Eq. (27) vary in $[-0.5, 4]$ with a step of 0.5.

In the following, three different analyses have been carried out:

(i) Single model analysis: elaboration of one random train/validation dataset

(ii) Statistical analysis: investigation on the effect of different train/validation datasets on the performance of CNNs and the CNN-DST

(iii) Noise analysis: investigation on the effect of noise in the data on the performance of the CNN-DST

(iv) Bandwidth analysis: investigation on the role of different bandwidth of the FRFs on the overall accuracy of the CNN-DST.

## 4.1 Single model analysis

In this section, one random train/validation dataset has been selected to highlight the steps of the CNN-DST framework. The accuracy of three sets of CNNs assessed on the validation dataset is reported in Table 2. They are (i) *All frequencies I*: CNNs trained by the |FRF| at all the frequencies, (ii) *All frequencies II*: CNNs trained by the |FRF| at all the frequencies. In this set, as proposed in [16], a large kernel size, i.e. 100, is selected for the first convolution layer, (iii) *Selected frequencies*: CNNs trained by the |FRF| at frequencies selected by the LARS approach to emphasize the importance of the frequency selection. Further, the required times for performing the frequency selection and training the CNNS are reported. As can be observed, the CNNs in *All frequencies I* and *II* have similar performances but the one with the smaller kernel size required less training time. More importantly, one can observe that by performing frequency selection ~85% of the total time was saved, and also the average accuracy was improved by ~1.5%. A closer look at the table reveals that the CNNs trained on the selected frequencies outperform the ones trained on the full frequency range. This implies that employing LARS as a data reduction/selection scheme, not only reduces the computational time but makes the information hidden in the frequencies more accessible for the CNNs too. From here on, only the CNNs trained on the selected frequencies are used as the constituent models.

The next step is to select proper CNNs for combination. In this regard, the CNNs have been ranked by using the proposed approach (see Section 3.2). The outcome is also shown in Table 2. Now, the first CNN was selected to initiate the ensemble. The other CNNs based on their rankings have been added to the ensemble one-by-one. The proposed fusion technique, see Algorithm 2, has been employed to combine their outputs. The accuracy of the ensemble over the validation dataset has been assessed by varying the DST parameter ($\theta$) in the range [-0.5, 4] with a step of 0.5. The obtained results are shown in Figure 5b. The figure has been color-coded based on the performance of the ensemble by changing the number of models and the $\theta$ value. Since the best constituent model has a validation accuracy of 95.65% (see Table 2), ensembles with an accuracy of 95.65% are shown in yellow, the accuracies larger than 95.65% in the green spectrum, and the accuracies lower than 95.65% in the red spectrum. Figure 5a shows the maximum accuracy that could be achieved by the available CNNs at any DST parameter and with the different numbers of models in the ensemble. This is obtained by an extensive search over the whole model space. It provides a baseline to demonstrate the efficacy of the proposed classifier selection methodology. As can be observed, the proposed classifier selection algorithm could reach the maximum possible accuracy provided that proper DST parameter ($\theta$) was selected. Another important point of consideration is the execution time of each approach. The proposed approach took 1.39 s whereas, the extensive search over the model space took 87.5 s. This means, by using the proposed classifier selection procedure, we could reach the maximum achievable accuracy about 60 times faster.

Table 2. Simulation time (in sec) and validation accuracies (in %) of the model trained by the |FRF| at the frequencies selected by the LARS approach (Selected frequencies), at all the frequencies with small kernel size (All frequencies I), and with large kernel size (All frequencies II).

|  | Selected frequencies | | | All frequencies I | | All frequencies II | |
| --- | --- | --- | --- | --- | --- | --- | --- |
|  | Time | Accuracy | rank | Time | Accuracy | Time | Accuracy |
| **Freq. Selection** | 67.12 | ---- | ---- | ---- | ---- | ---- | ---- |
| **CNN($X_1$)** | 103.08 | 92.75 | 7 | 625.08 | 92.75 | 871.02 | 89.90 |
| **CNN($X_2$)** | 94.99 | 95.65 | 9 | 616.06 | 94.20 | 874.41 | 92.75 |
| **CNN($X_1 + X_2$)** | 95.24 | 94.20 | 3 | 609.25 | 92.75 | 920.98 | 94.20 |
| **CNN($X_1 \circ X_2$)** | 93.56 | 95.65 | 1 | 603.50 | 94.20 | 955.53 | 92.75 |
| **CNN($X_1^2$)** | 94.89 | 91.30 | 4 | 605.84 | 87.09 | 887.97 | 87.09 |
| **CNN($X_2^2$)** | 91.23 | 94.20 | 5 | 600.83 | 94.20 | 867.73 | 92.75 |
| **CNN($\log(X_1)$)** | 95.31 | 94.20 | 2 | 600.14 | 92.75 | 882.14 | 94.20 |
| **CNN($\log(X_2)$)** | 92.13 | 92.75 | 8 | 630.96 | 89.90 | 952.24 | 91.30 |
| **CNN($All$)** | 335.50 | 95.65 | 6 | 3050.76 | 95.65 | 4542.47 | 94.20 |
| **Total** | 1163.05 | 94.04* | ---- | 7942.42 | 92.61* | 11754.49 | 92.13* |

\* This is the average of the accuracies

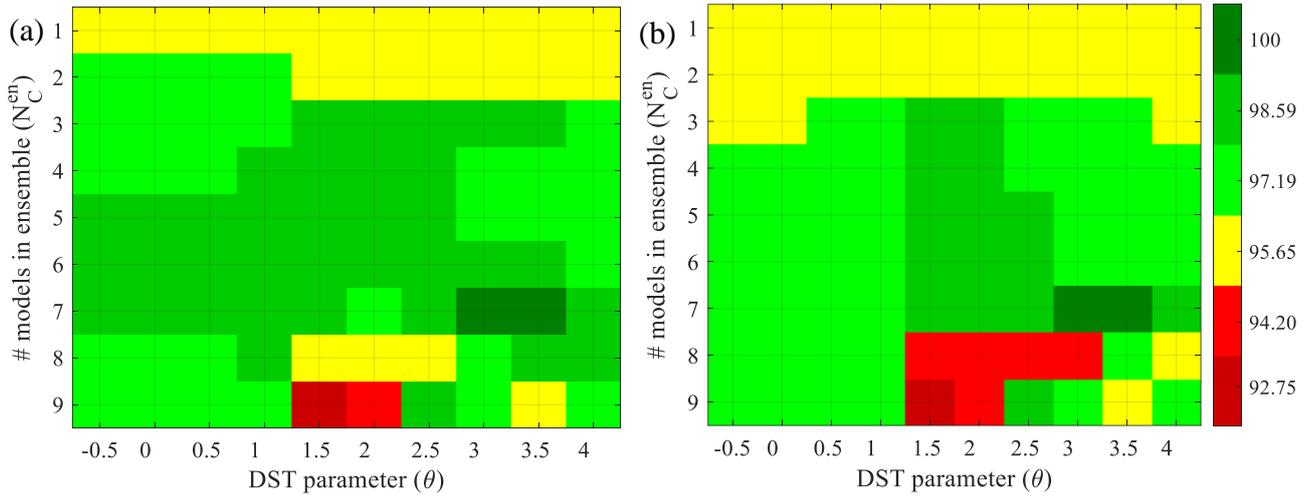

Figure 5. Analyzing the effect of the number of models in the ensemble, and DST parameter ($\theta$) on the validation accuracy of the ensemble. (a) maximum achievable accuracy by the proposed fusion method, obtained by an extensive search over the model space, (b) accuracy achieved by employing the proposed fusion method together with the proposed classifier ranking/selection procedure.

## 4.2 Statistical analysis

In this section, a statistical analysis has been performed to provide a detailed discussion on the performance of the CNN-DST. For this purpose, the whole procedure, i.e. from data resampling to the fusion output, has been repeated 50 times. The variations in the validation accuracies of the CNNs are shown in the form of the violin plot in Figure 6. The shapes of the violins indicate the distribution of the accuracies. Their mean and median locations (resp. values) are shown in the black and red lines (resp. numbers), respectively. It can be seen that the CNNs trained by the logarithm of the |FRF|s give the best performance. This could emphasize the effect of anti-resonances in detecting the defected samples. In other words, the logarithm of the FRF reduces its dynamic range, and thus the contribution of the anti-resonances to detect faulty samples becomes more pronounced. That means the information in the anti-resonances becomes more accessible to be used for training the CNNs. This could justify the higher accuracy of the CNNs trained by the $\log(X)$.

Further, it can be seen that the proposed CNN-DST framework outperforms the constituent models. The mean, respectively median, of the constituent models, are in the range of 92.32%-94.72%, 92.75%-95.65% respectively. The CNN-DST framework provides an increased mean and median of 97.19% and 97.10% respectively, and as such outperforms the best constituent model.

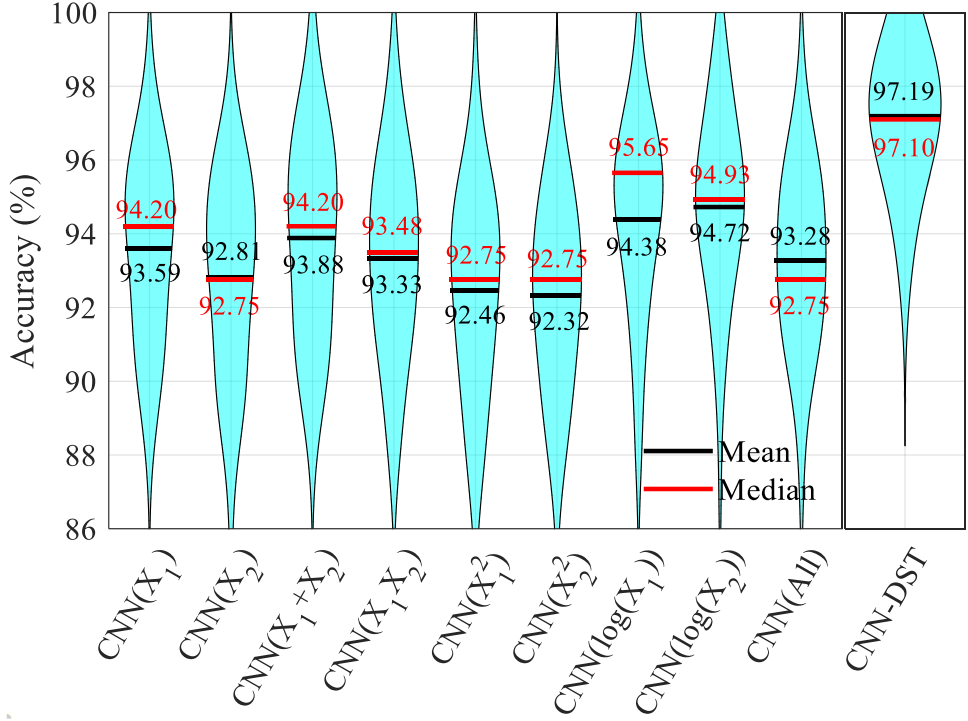

Figure 6. Variation in the validation accuracy of the CNNs and the CNN-DST. The results were obtained through 50 iterations. Means and medians are shown respectively in black and red lines and numbers.

Figure 7 provides a comparative analysis of the accuracy of the CNN-DST frameworks obtained by different ranking and fusion techniques. In this figure, the proposed ranking and fusion methods are indicated respectively by $r_p$ and $F_p$. $r_l$ is a classifier ranking strategy from the literature [59]. $F_l$ is a fusion method proposed by the authors. It is similar to $F_p$ but without considering the support degree to the chief focal element mentioned in step 7 of Section 3.3, i.e. when the DST parameter $\theta = 1$ [38]. The utilized methods for classifier ranking and fusion method were explicitly mentioned in the abscissa of Figure 7. For instance, the label $\{r_p, F_l\}$ indicates that the ranking is based on the proposed ranking $r_p$, while the fusion is based on the fusion method $F_l$.

$\{F_l\}$ and $\{F_p\}$ shows the maximum accuracy that could be achieved by each method. They have been obtained by an extensive search over the whole model space. The figure indicates that both fusion methods $\{F_p\}$ and $\{F_l\}$ could outperform the constituent models, with a better performance of the here proposed fusion method $F_p$.

The figure indicates that the procedure from literature, $\{r_l, F_l\}$, could outperform the individual classifiers. Changing the classifier ranking from $r_l$ to the proposed $r_p$, i.e. $\{r_p, F_l\}$ could improve the classification accuracy. Now, by changing the fusion method from $F_l$ to $F_p$, a significant improvement on the classification performance has occurred, from 96.03% for $\{r_l, F_l\}$ to 97.10% for $\{r_l, F_p\}$. Using both the proposed ranking strategy and proposed fusion approach, i.e. $\{r_p, F_p\}$, yields the best classification performance with a mean and median value of 97.19% and 97.10% respectively. Comparing these results with their associated fusion references, i.e. $\{F_l\}$ and $\{F_p\}$, reveals that both classifier ranking algorithms are suboptimal but the proposed one gives the ranking closer to the optimal results.

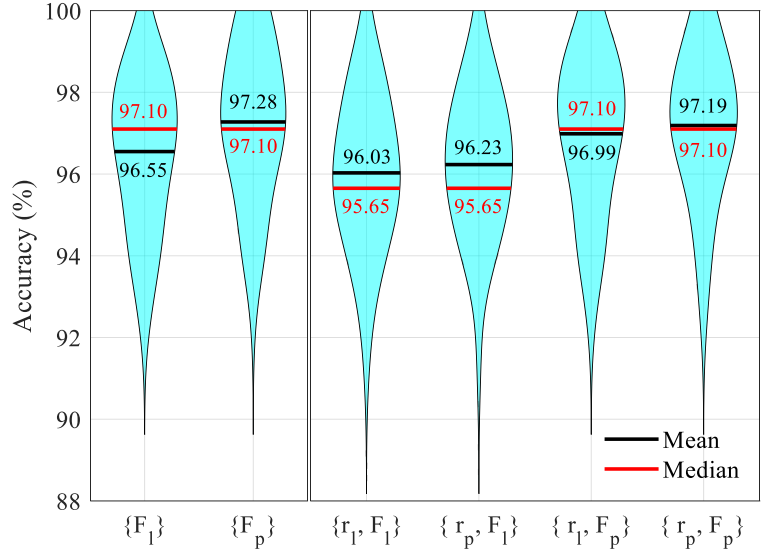

Figure 7. Variation in the validation accuracy of the CNN-DSTs obtained by using the proposed classifier ranking $r_p$ and fusion $F_P$ methods together with the one from the literature $r_l$ and $F_l$. The results were obtained through 50 iterations. Means and medians are shown respectively in black and red lines and numbers.

## 4.3 Noise analysis

This section investigates the robustness of the proposed CNN-DST method in presence of different noise levels. For this purpose, random Gaussian noises with Root-Mean-Square (RMS) 10, 20, 50, 80, 120, and 160% of the signal RMS have been added to the measured FRF. For each case, the Signal-to-Noise ratio (SNR) has been estimated by using the one-channel-based method [60,61] and shown in the form of a boxplot in Figure 8. In this figure, red lines and black circles indicate medians and mean values. It can be observed that the measured FRF has an average SNR of 77 dB, then by adding 10, 20, 50, 80, 120, 160% rms NSR white Gaussian noise to the signal its quality has been reduced respectively to average SNR of 40, 29, 14, 6.5, 1.5, and -2.5 dB. Figure 9 illustrates a random FRF with SNR = 71.08 dB and when the white Gaussian noise with RMS equal to 160% of the FRF RMS has been added to it. As can be seen, adding such noise to the signal reduced the quality of the signal such that its SNR has been dropped to -0.80 dB.

At each noise level, the whole procedure of the CNN-DST framework, as shown in Figure 1, have been carried out 50 times for statistical interpretation. Variations in the validation accuracy of the individual CNNs have been presented in Appendix II. Figure 10 shows the effect of noise on the validation accuracy of the proposed CNN-DST framework. Their means and medians are shown in the black and red lines.

One can observe that by adding low levels of noise to the FRF, up to 20% RMS, the average accuracy has been reduced from 97.19% to 96.32%. Since these noise-levels mostly bury the information in the low-amplitude frequencies, e.g. anti-resonances, the importance of such frequencies for classification can be inferred. The reason for the drop of median accuracy in 10% RMS noise is twofold: few numbers of iterations in the statistical analysis, and few numbers of samples in the validation dataset. Another accuracy reduction occurs by adding very high noise-levels, above 80% RMS, to the FRF. This is due to the fact that these high noise-levels hide the information in some of the peaks in the FRF, see Figure 9. Therefore, it implies the importance of the peak frequencies for classification. Another interesting observation here is that no changes in the accuracy happened in the noise-levels between 20%-80% RMS. It indicates that the frequencies buried in these noise-levels, which are mostly neither resonance nor anti-resonance, carry very limited information. It is in accordance with the basics of vibration theory and modal analysis theory [62].

Moreover, by comparing the distribution of the accuracies and their associated mean and medians in presence of different noise levels, up to 160% RMS NSR, one can see that even though the accuracy of constituent CNNs have been reduced, see Appendix II, the proposed ensemble method CNN-DST remains accurate. By that, the robustness of the CNN-DST with respect to the measurement noise can be inferred.

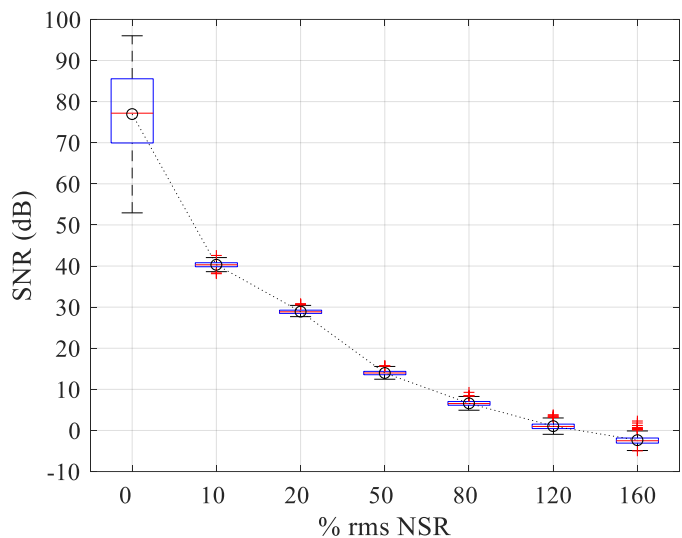

Figure 8. Signal-to-Noise ratio (SNR) of the input FRF polluted with additive white Gaussian noise with different RMS Noise-to-Signal ratios (NSR). The red lines and black circles show the medians and mean-values, respectively.

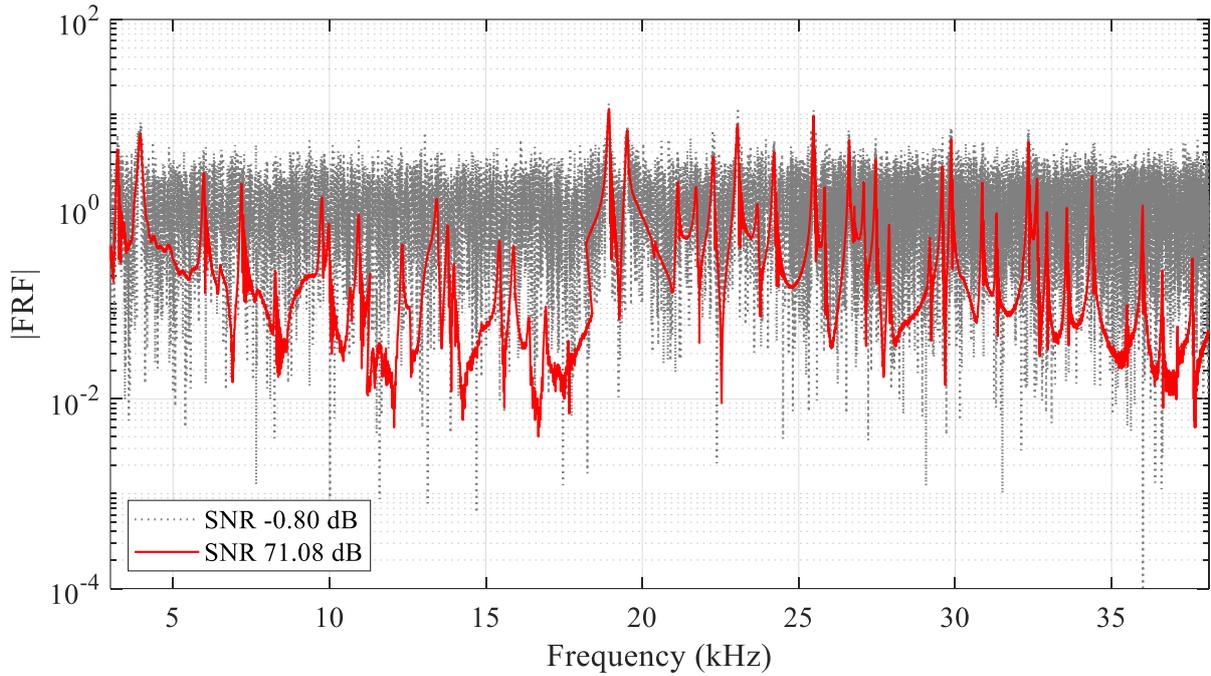

Figure 9. Comparison between a measured FRF with SNR=71.08 dB and when it is polluted with white noise with RMS 160% RMS of the FRF signal. It is equivalent to SNR=-0.80 dB.

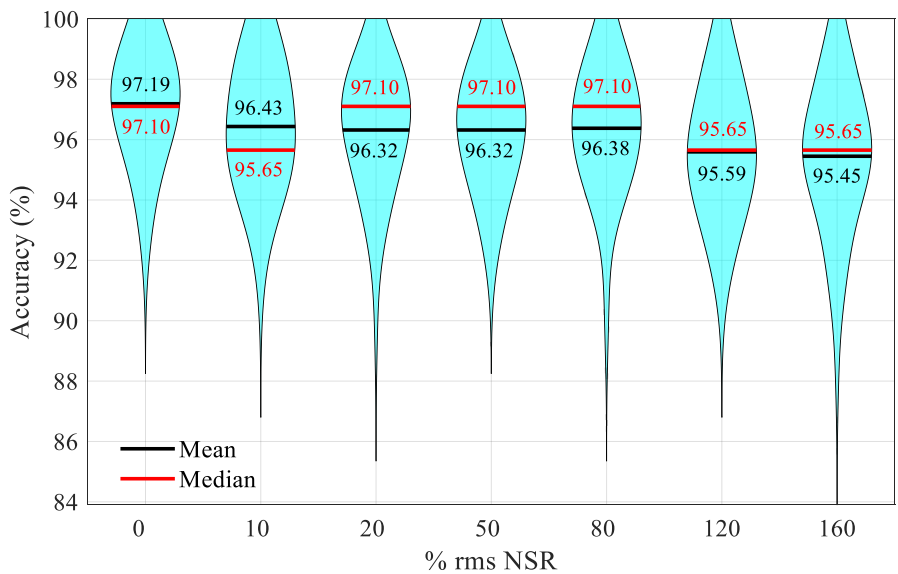

Figure 10. Variation in the validation accuracy of the CNN-DST framework in presence of different levels of noise. The results were obtained through 50 iterations. Means and medians are shown respectively in black and red.

### 4.4 Bandwidth analysis

Ideally, to reduce the time and complexity of the experiments, a small frequency range of the FRFs is seeking which contains most of the required information for detecting faulty samples. In this section, the role of different bandwidths of the FRF on the classification performance is investigated. For this purpose, two cases have been investigated, Case I: the measured FRFs and Case II: the measured FRFs polluted with the addition of 50% RMS noise. One sample FRF of each case is shown in Figure 11.

For each case, the frequency range of the FRFs has been split into 2, 4, 8, 16 equal sections and their associated CNN-DST has been obtained through 10 iterations. The results are presented in Table 3 and Table 4 for Case I and Case II respectively. In the tables, for each case and split scenarios, the most accurate section of the FRF is bolded. The following statements can be inferred:

(i) By reducing the frequency range, some information might be lost. However, by selecting the proper section of the FRF, the lost information can be minimized. For instance, by splitting the frequency range into 16 sections, there is one section in each case that by using its information, the classification accuracy has been reduced less than 1%. In Figure 11, these sections have been shown in red and grey rectangles for Case I and Case II, respectively.

(ii) In both cases, most of the essential information for classification is available in the first half of the FRF. This can be interpreted as, for the here studied cases, the peaks in the first half are affected most by the presence of defects.

(iii) By splitting the FRFs into 8 equal sections, the maximum accuracy occurs in the second and first sections respectively, for Case I and Case II. The averaged accuracy in the first and fourth sections of Case II is equal to those of Case I whereas, the accuracy in the other sections have been dropped significantly. This implies that the peaks that are above the noise level in these regions convey vital information for classification. But in other sections, the important peaks have been buried in the noise

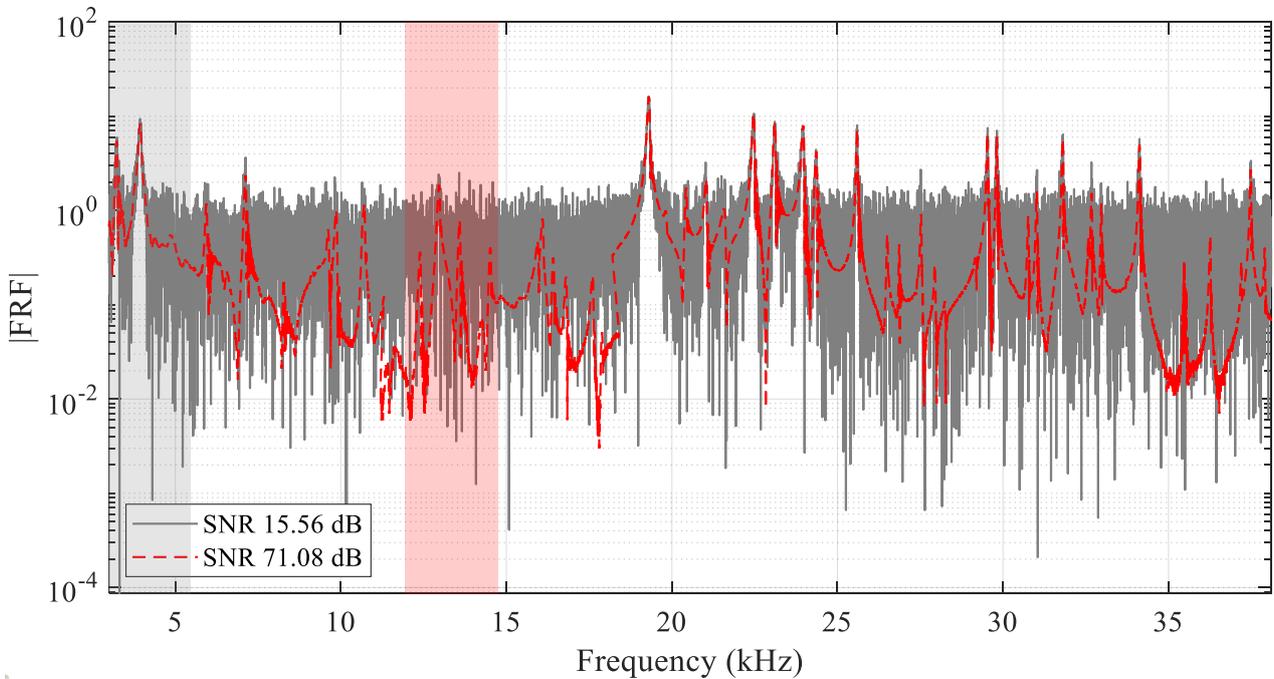

Figure 11. One sample FRF with SNR=71.08 dB (Case I) and when it is polluted with 50% root-mean-square noise-to-signal ratio white noise (Case II). It is equivalent to SNR=15.56 dB. The red and grey rectangles respectively indicate the most accurate portion of the measured FRF and the noisy FRF.

(iv) By splitting the FRF into 16 sections, the maximum accuracy occurs in the fourth and first sections respectively, for Case I and Case II. This implies that the most important peaks for our purpose are the ones in the fourth and first sections. For Case II, this peak has been partially buried in the noise and thus, some part of the accuracy has been lost.

# 5 Conclusion

An ensemble deep learning framework based on the convolutional neural network CNN and Dempster-Shafer theory DST has been developed for vibration-based fault recognition. The proposed CNN-DST framework consists of three main stages: (i) CNN generation, (ii) information-based CNN selection and (iii) DST-based CNN fusion. Moreover, employing a frequency selection scheme prior to training the CNNs was also proposed. It has been applied to the quality classification of polycrystalline Nickel alloy first-stage turbine blades. For this, an experimental dataset was created by conducting vibrational broadband measurement over 229 turbine blades with different types and severities of damages, and the |FRF| data is fed to the CNN-DST framework. It has been shown that employing the frequency reduction/selection scheme prior to training the CNNs could drastically reduce the computational time without losing its accuracy. A statistical analysis reveals that the average classification accuracy of the ensemble CNN-DST equals 97.19%. The CNN-DST outperforms the individual constituent CNN model as well as a state-of-the-art fusion method in terms of classification performance for the here discussed test case of turbine blades. Moreover, the recorded |FRF| data was synthetically polluted with additive white Gaussian noise of different levels, up to 160% RMS NSR which corresponds to the average SNR of -2.5 dB, and their effects on the accuracy of the CNN-DST have been statistically investigated. The results illustrated the robustness of the proposed framework against noise. Finally, a bandwidth analysis indicated that by reducing the frequency range of the FRF by a factor of 16, less than 1% of the classification performance was lost. Though, this is on the condition that the proper frequency range in the FRF is chosen.

As was shown, in some cases the validation accuracy could reach 100% accuracy. This means that some samples and/or frequencies contain more valuable information for training. Quantifying the information content of the samples/frequencies and the importance of their presence in the training dataset is being sought by the authors as a future study.

Table 3. Accuracy analysis of the CNN-DST framework by using one portion out of 2, 4, 8, 16 portions of the measured FRF. The first row indicates the first frequency of each portion. The results were obtained through ten iterations.

| Freq (kHz) | 3.00 | 5.91 | 8.81 | 11.62 | 14.59 | 17.56 | 20.46 | 23.61 | 25.70 | 27.31 | 28.92 | 30.47 | 32.12 | 33.71 | 35.41 | 36.71 |
|---|---|---|---|---|---|---|---|---|---|---|---|---|---|---|---|---|
| Accuracy (%) | **97.19 ± 1.79** ||||||||||||||||
|  | **96.73 ± 1.96** |||||||| 95.22 ± 1.19 ||||||||
|  | **96.23 ± 1.50** |||| 95.36 ± 2.03 |||| 93.48 ± 2.18 |||| 92.75 ± 2.65 ||||
|  | 95.65 ± 1.70 || **96.23 ± 1.83** || 95.80 ± 2.93 || 95.50 ± 2.31 || 89.57 ± 3.67 || 88.99 ± 1.40 || 91.45 ± 2.20 || 79.86 ± 3.95 ||
|  | 95.65 ± 2.37 | 93.33 ± 1.83 | 94.93 ± 1.23 | **96.23 ± 1.96** | 93.33 ± 2.39 | 95.80 ± 2.41 | 95.07 ± 3.22 | 92.03 ± 2.29 | 80.87 ± 2.87 | 85.65 ± 4.24 | 91.45 ± 3.16 | 78.84 ± 4.80 | 79.86 ± 3.52 | 90.00 ± 2.10 | 72.61 ± 4.50 | 79.85 ± 4.18 |

Table 4. Accuracy analysis of the CNN-DST framework by using one portion out of 2, 4, 8, 16 portions of the measured FRF polluted with 50% RMS NSR white Gaussian noise. The first row indicates the first frequency of each portion. The results were obtained through ten iterations.

| Freq (kHz) | 3.00 | 5.91 | 8.81 | 11.62 | 14.59 | 17.56 | 20.46 | 23.61 | 25.70 | 27.31 | 28.92 | 30.47 | 32.12 | 33.71 | 35.41 | 36.71 |
|---|---|---|---|---|---|---|---|---|---|---|---|---|---|---|---|---|
| Accuracy (%) | **96.32 ± 1.92** ||||||||||||||||
|  | **96.23 ± 2.07** |||||||| 94.20 ± 2.16 ||||||||
|  | **95.65 ± 2.27** |||| 95.22 ± 1.68 |||| 90.72 ± 2.99 |||| 90.29 ± 2.73 ||||
|  | **95.65 ± 2.98** || 94.20 ± 2.16 || 91.88 ± 1.01 || 95.51 ± 1.59 || 81.30 ± 2.69 || 87.54 ± 3.70 || 89.28 ± 3.70 || 81.45 ± 4.31 ||
|  | **95.65 ± 2.37** | 94.93 ± 2.92 | 94.63 ± 2.56 | 86.37 ± 3.94 | 94.20 ± 3.61 | 88.55 ± 3.01 | 93.62 ± 1.83 | 88.41 ± 3.13 | 76.80 ± 3.98 | 82.90 ± 5.15 | 89.86 ± 4.16 | 72.61 ± 2.31 | 76.10 ± 2.39 | 88.84 ± 2.74 | 72.75 ± 2.25 | 82.17 ± 2.74 |

# 6  Acknowledgment

The authors gratefully acknowledge the ICON project DETECT-ION (HBC.2017.0603) which fits in the SIM research program MacroModelMat (M3) coordinated by Siemens (Siemens Digital Industries Software, Belgium) and funded by SIM (Strategic Initiative Materials in Flanders) and VLAIO (Flemish government agency Flanders Innovation & Entrepreneurship). Vibrant Corporation is also gratefully acknowledged for providing anonymous datasets of the blades.

# Appendix I.  Example on the fusion method

Suppose there are four individual classifiers ($C_1$, $C_2$, $C_3$, and $C_4$) trained on a three classification problem. Given an instance $X$, these classifiers give the output vector $\hat{y}_1 = (0.5, 0.1, 0.4), \hat{y}_2 = (0.3, 0.3, 0.4), \hat{y}_3 = (0.5, 0.0, 0.5)$, and $\hat{y}_4 = (0.4, 0.2, 0.4)$. Here $n_c = 3$, $N_c = 4$. Therefore, $\mathfrak{E} = \{E_1, E_2, E_3\}$, $m_i = \hat{y}_i$ and $m_i(\mathfrak{E}) = 0$. To combine the classifiers based on the proposed fusion technique, the following steps should be taken:

Step 1. To obtain the average BJS of the BOEs, take $m_1$ and $m_i, i = 1, \dots, 4$, $BJS(m_1, m_i) = [0.0, 0.056, 0.054, 0.0163]$ thus $aBJS_1 = 0.042$. Do the procedure for $m_i$, $i = 1, \dots, 4$ leads to $aBJS = [0.042, 0.080, 0.111, 0.046]$.

Step 2. The Disagreement associated with each BOEs are obtained as follows,

I. $\bar{m} = [0.425, 0.150, 0.425]$, $d_J(m_i, \bar{m}) = [0.094, 0.197, 0.184, 0.061]$ and thus $SW = 0.134$.

II. Take $m_1$, $\bar{m}_{\sim 1} = [0.400, 0.167, 0.433]$ and $d_J(m_i, \bar{m}_{\sim 1}) = [0.170, 0.205, 0.047]$, and thus $SW_{\sim 1} = 0.0141$. Do the procedure for $m_i$, $i = 1, \dots, 4$ leads to $SW_{\sim i} = [0.141, 0.099, 0.094, 0.154]$.

III. $m^*_{\Delta E, \sim i} = [0.496, 0.522, 0.525, 0.487]$

Step 3. The support degree of the BOEs are

$$SD = [47.95, 23.87, 17.09, 45.02]$$

Step 4. Normalized support degree is

$$\widehat{SD} = [0.358, 0.178, 0.128, 0.336]$$

Step 5. The credibility of the BOEs are,

    I.    $E_d = [0.410, 0.473, 0.301, 0.458]$

    II.    $CD = [0.540, 0.286, 0.172, 0.532]$

Step 6. Normalized credibility is,

$$\widehat{CD} = [1.00, 0.53, 0.32, 0.99]$$

Step 7. support degree to the chief focal element

    I.    $\bar{m} = [0.425, 0.150, 0.425] \rightarrow E^{chief} = 1$

    II.    $SD^{chief} = [0.5, 0.3, 0.5, 0.4]$

    III.    $\widehat{SD}^{chief} = [1.00, 0.60, 1.00, 0.80]$

Step 8. Evaluate the weight

$$W = \theta[1.00, 0.53, 0.32, 0.99] + (1-\theta)[1.00, 0.60, 1.00, 0.80]$$

$$W = [1, 0.6 - 0.7\theta, 1 - 0.66\theta, 0.8 + 0.19\theta]$$

assuming $\theta = 0.5$, $W = [1.00, 0.56, 0.67, 0.89]$

step 9. Normalize the weight

$$W = [0.32, 0.18, 0.21, 0.29]$$

Step 10. Evaluate the weighted average evidences (WAE)

$$WAE = [0.44, 0.14, 0.42]$$

Step 11. The final output obtained by Demspter's rule of combination is,

$$\tilde{Y} = [0.83, 0.12, 0.05]$$

MaxIndex([0.83, 0.12, 0.05]) = 1, Hence, the instance $X$ belongs to class 1.

To investigate the effect of $\theta$ on predicting the class scores, it has been changed in the range of [-2, 2] and scores has been obtained. It is shown in Figure I-1. It can be seen that the fusion method could increase the score of class 1 to about 0.89.

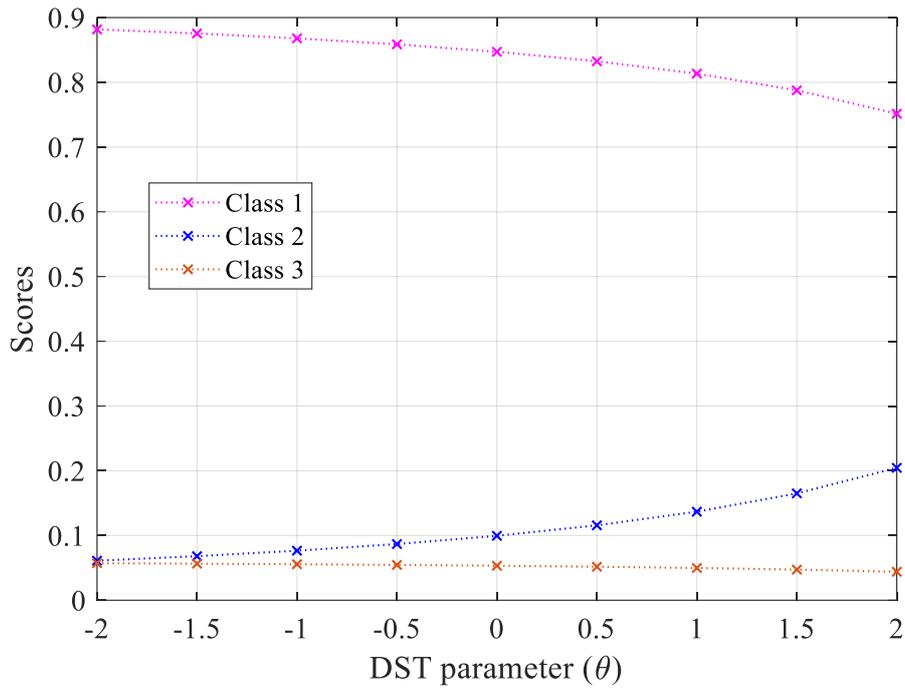

Figure I-1. The effect of DST parameter $\theta$ on the predicted scores.

# Appendix II.   CNN accuracies with different noise levels

In this appendix, the variation in the validation accuracies of the CNNs in presence of different noise levels are shown in the form of violin plots. Moreover, the validation accuracy of their associated CNN-DST is shown. The accuracy improvement by the CNN-DST can be simply observed.

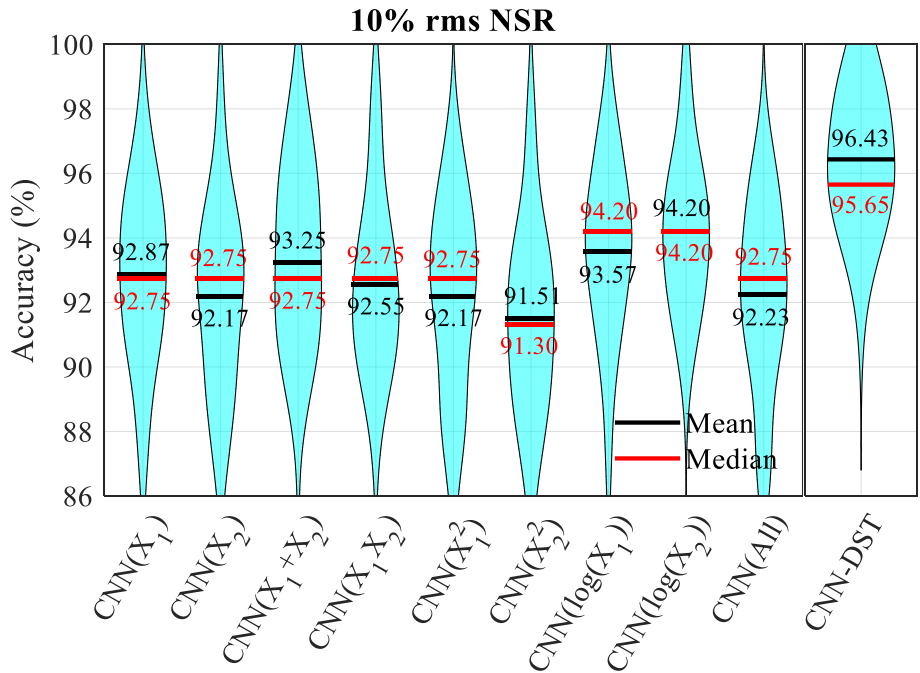

Figure II-1. Variation in the validation accuracy of the CNNs and their associated CNN-DST when the |FRF| had been polluted with 10%rms NSR. The results were obtained through 50 iterations. Means and medians are shown respectively in black and red lines and numbers.

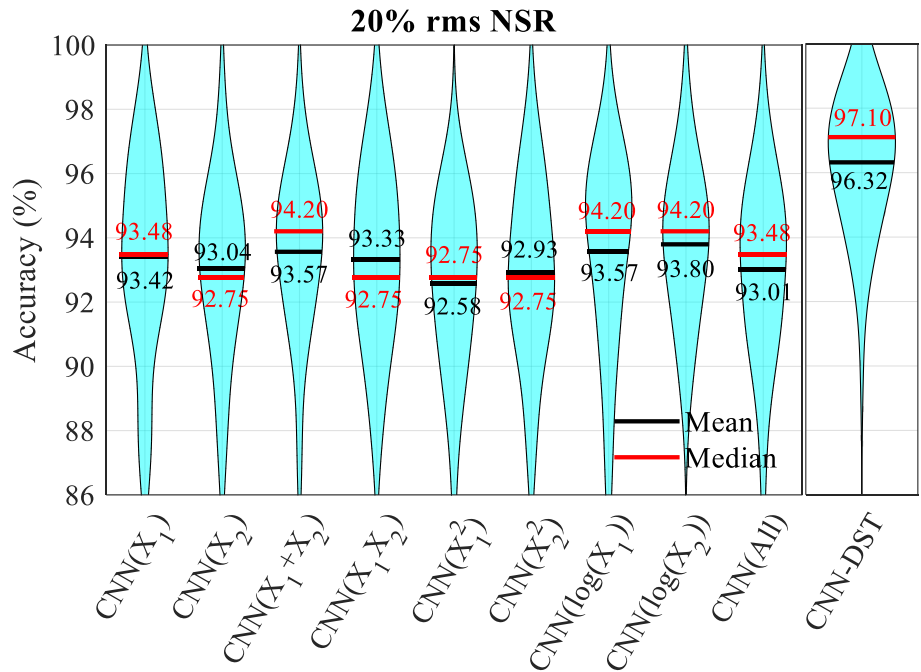

Figure II-2. Variation in the validation accuracy of the CNNs and their associated CNN-DST when the |FRF| had been polluted with 20%rms NSR. The results were obtained through 50 iterations. Means and medians are shown respectively in black and red lines and numbers.

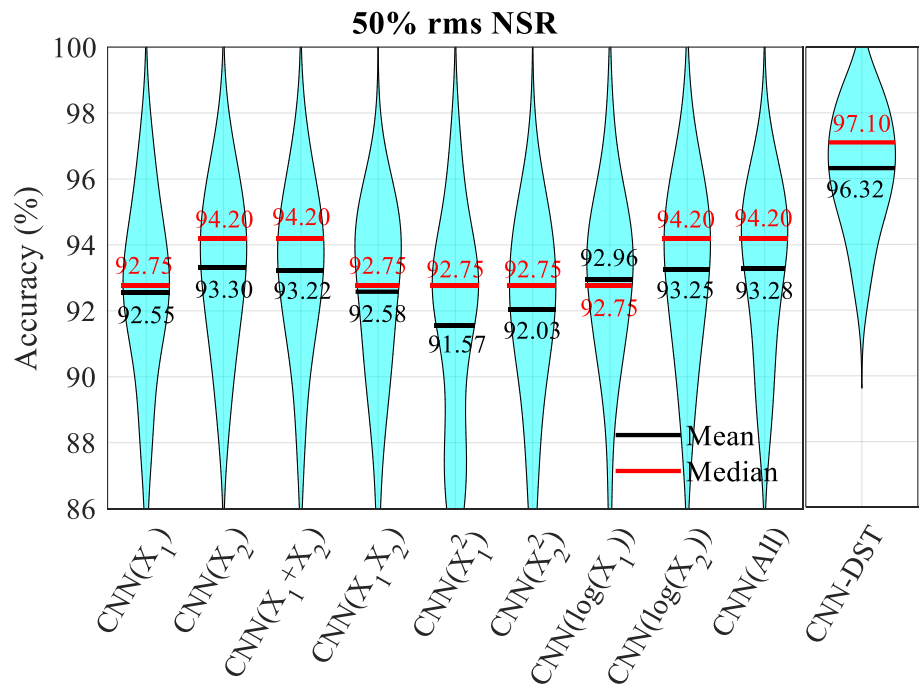

Figure II-3. Variation in the validation accuracy of the CNNs and their associated CNN-DST when the |FRF| had been polluted with 50%rms NSR. The results were obtained through 50 iterations. Means and medians are shown respectively in black and red lines and numbers.

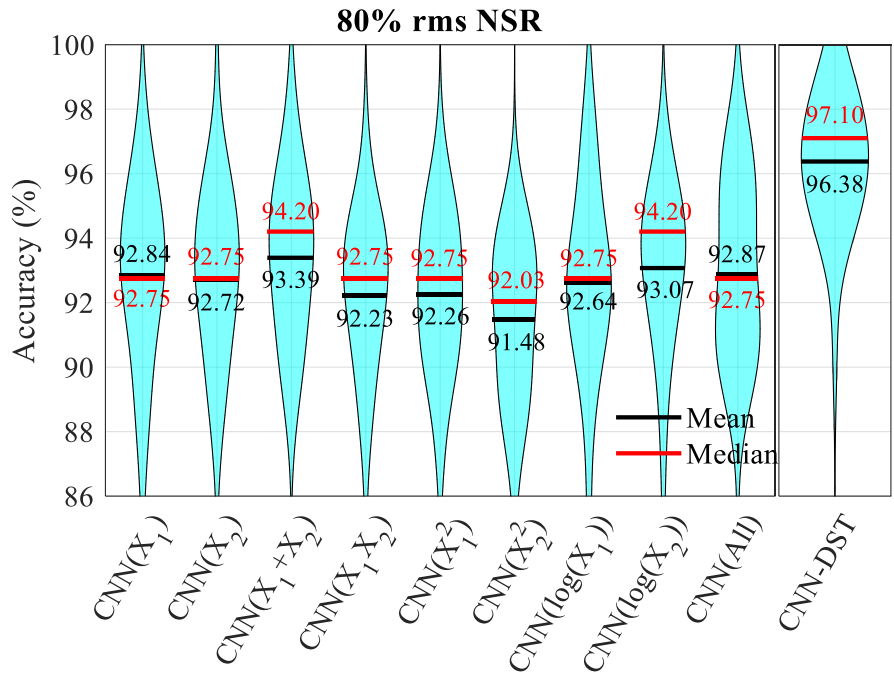

Figure II-4. Variation in the validation accuracy of the CNNs and their associated CNN-DST when the |FRF| had been polluted with 80%rms NSR. The results were obtained through 50 iterations. Means and medians are shown respectively in black and red lines and numbers.

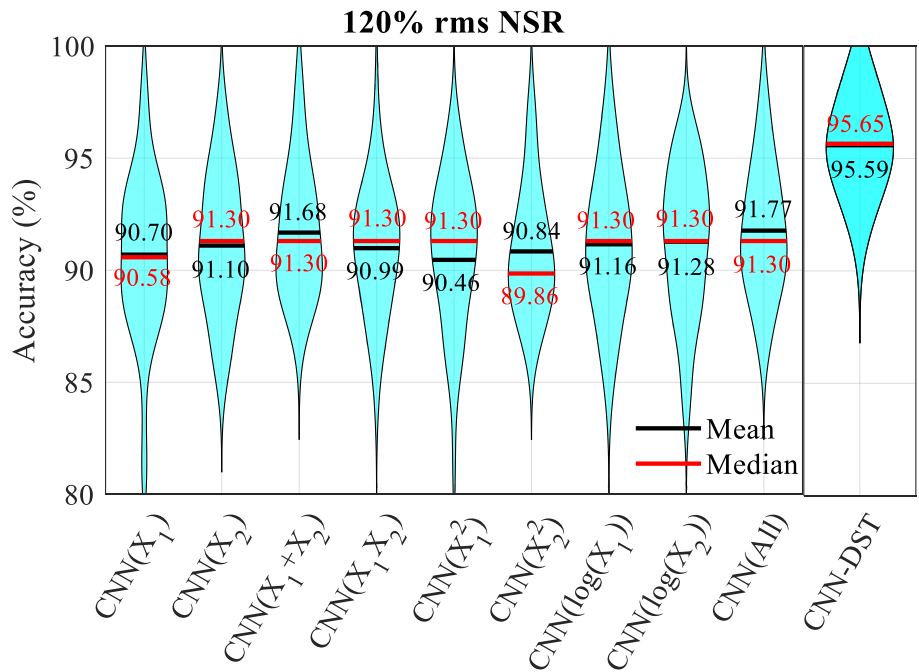

Figure II-5. Variation in the validation accuracy of the CNNs and their associated CNN-DST when the |FRF| had been polluted with 120%rms NSR. The results were obtained through 50 iterations. Means and medians are shown respectively in black and red lines and numbers.

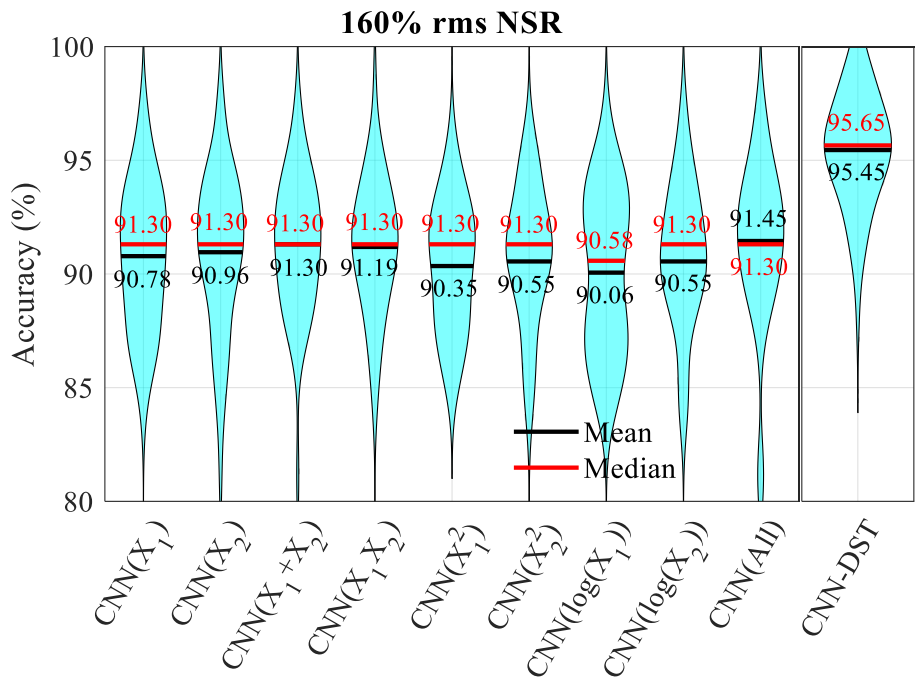

Figure II-6. Variation in the validation accuracy of the CNNs and their associated CNN-DST when the |FRF| had been polluted with 160%rms NSR. The results were obtained through 50 iterations. Means and medians are shown respectively in black and red lines and numbers.